\documentclass{emulateapj}
\usepackage{graphicx,amsmath,natbib,bm}

\usepackage{epsf}
\usepackage{epstopdf}

\input{epsf}
\usepackage{epsfig}

\usepackage{amsfonts}
\DeclareGraphicsExtensions{.jpg,.pdf,.png,.eps,.ps}

\newcommand{\as}{\ensuremath{^{\prime\prime}}}

\def\msol   {\ifmmode{{\rm M}_{\odot} }\else{M$_{\odot}$}\fi}
\def\lsol   {\ifmmode{{\rm L}_{\odot}}\else{${\rm L}_{\odot}$}\fi}
\def\cii    {\ifmmode{{\rm C}{\rm \small II}}\else{C{\scriptsize II}}\fi}

\slugcomment{}
\shorttitle{DM Substructure Detection with ALMA}
\shortauthors{Hezaveh et al.}
\begin{document}

\title{Dark Matter Substructure Detection Using Spatially Resolved
  Spectroscopy of Lensed Dusty Galaxies}
\author{Yashar Hezaveh\altaffilmark{1}, Neal Dalal\altaffilmark{2},
  Gilbert Holder\altaffilmark{1}, Michael Kuhlen\altaffilmark{3}, 
  Daniel Marrone\altaffilmark{4}, Norman Murray\altaffilmark{5,6}, Joaquin Vieira\altaffilmark{7}}
\altaffiltext{1}{Department of Physics,
McGill University, 3600 Rue University, 
Montreal, Quebec H3A 2T8, Canada}
\altaffiltext{2}{Astronomy Department, University of Illinois at
  Urbana-Champaign, 1002 W.\ Green Street, Urbana IL 61801}
\altaffiltext{3}{Theoretical Astrophysics Center, University of California, Berkeley, CA 94720}
\altaffiltext{4}{Steward Observatory, University of Arizona, 933 North Cherry Avenue, Tucson, AZ 85721, USA}
\altaffiltext{5}{CITA, University of Toronto, 60 St.\ George St.,
  Toronto ON M5S 3H8, Canada}
\altaffiltext{6}{Canada Research Chair in Astrophysics}
\altaffiltext{7}{California Institute of Technology, 1200 East California Blvd, MC 249-17, Pasadena, CA 91125, USA}

\begin{abstract}  
We investigate how strong lensing of dusty, star-forming galaxies by
foreground galaxies can be used as a probe of dark matter halo
substructure.  We find that spatially resolved spectroscopy of
lensed sources allows dramatic improvements to measurements of lens
parameters.  In particular we find that modeling of the full,
three-dimensional (angular position and radial velocity) data can
significantly facilitate substructure detection, increasing the
sensitivity of observables to lower mass subhalos. We carry out
simulations of lensed dusty sources observed by early ALMA (Cycle 1)
and use a Fisher matrix analysis to study the parameter degeneracies
and mass detection limits of this method.  We find that, even with
conservative assumptions, it is possible to detect galactic dark
matter subhalos of $\sim 10^8\, M_{\odot}$ with high significance in
most lensed DSFGs.  Specifically, we find that in typical DSFG
lenses, there is a $\sim 55\%$ probability of detecting a
substructure with $M>10^8 \, M_{\odot}$ with more than $5 \sigma$
detection significance in {\em each lens}, if the abundance of
substructure is consistent with previous lensing results.  The full
ALMA array, with its significantly enhanced sensitivity and
resolution, should improve these estimates considerably.  Given the
sample of $\sim 100$ lenses provided by surveys like the South Pole
Telescope, our understanding of dark matter substructure in typical
galaxy halos is poised to improve dramatically over the next few
years.
\end{abstract}

\keywords{dark matter ---
gravitational lensing ---
galaxies: dwarf ---
galaxies: luminosity function, mass function ---
galaxies: structure ---}

\section{introduction}

The inflationary $\Lambda$CDM model has proven spectacularly
successful in describing the observed large-scale structure of the
universe.  Precise measurements of the mean expansion history of the
universe \citep[e.g.][]{Suzuki12}, the anisotropies of the cosmic microwave
background \citep{WMAP7}, and the clustering of galaxies at low
redshift $z\lesssim 1$ \citep[e.g.][]{BOSS_RSD,BOSS_cosmo} are all
consistent with a simple scenario involving a nearly scale-invariant
spectrum of curvature perturbations generated during inflation, whose
growth over cosmic time is governed by gravitational instability.

While measurements of large-scale ($\gtrsim 10$ Mpc) structure have
yielded the most stringent constraints on this cosmological model,
there is considerable information to be gleaned from observations of
structure on smaller scales as well.  For example, the detailed shape
of the inflationary potential influences the shape of the primordial
power spectrum \citep[e.g.][]{Dodelson03}, motivating efforts to
measure the scalar spectral index $n_s$ and its running 
$dn_s/d\log k$.  For example, a sharp feature in the inflationary
potential will generate features in the primordial power spectrum,
breaking its near scale-invariance \citep{Kamion00}.
In addition, the particle physics of dark matter
affects the shape of the matter transfer functions, leading to a
Jeans-like suppression of structure below the free-streaming scale of
the dark matter particles.  In WIMP-like scenarios, this
free-streaming scale is as small as 1 comoving pc \citep{Loeb05},
however in alternative dark matter models this damping scale can be
much larger 
\citep[e.g.][]{Bode01,Abazajian06,Cembranos05,Kaplinghat05,CyrRacine12}.  
Accordingly,
measurements of the small-scale power spectrum can in principle
provide a wealth of information about the physics of the early universe.

To date, the most constraining bounds on the small-scale power
spectrum have been derived from observations of the Lyman-$\alpha$
forest \citep{Seljak06}.  However, it will be difficult for future LyAF
observations to improve significantly on existing bounds because
current, high-resolution spectra already have resolution approaching
the Jeans scale of the intergalactic medium at $z\sim 2-3$.  Another
probe of small-scale structure is the abundance of low-mass dark
matter halos and subhalos.  High-resolution N-body simulations
\citep{ViaLactea1,ViaLactea2,GHalo,Aquarius} have revealed that dark
matter halos in hierarchical CDM cosmologies are expected to contain
copious substructure, primarily in the form of gravitationally
self-bound, dynamically cold subhalos with a spectrum of masses.  The
abundance of these subhalos is expected to depend on the amplitude and
shape of the small-scale power spectrum \citep[e.g][]{Zentner03},
meaning that measurements of halo substructure should constrain
early-universe physics.

Numerous groups have attempted to constrain the abundance of low-mass
subhalos by detecting faint, DM-dominated dwarf satellite galaxies in
the Local Group \citep[see][for a review]{Kravtsov10}.  Despite heroic
efforts, the number of detected dwarf satellites falls well below the
expected abundance of low-mass subhalos for a typical galactic halo.
This discrepancy has been termed the ``Satellite Problem'' for CDM
models, and has led to considerable speculation that the physics of
dark matter might not be described well by the idealized,
non-interacting, absolutely cold CDM model.  However, astrophysical
processes can plausibly suppress the star-formation
efficiency of low-mass halos and subhalos, rendering many of
them invisible to optical surveys. To distinguish between
astrophysical solutions, and particle-physics solutions to the
Satellite Problem, a purely gravitational method for detecting
potentially dark subhalos is required.  One possible means of
detecting dark substructure in galactic halos is to search for its
dynamical effects on cold stellar streams \citep{Carlberg11,Yoon11}.
Another gravitational probe
of dark matter substructure is strong gravitational lensing, which is the focus of this work.

\citet{MaoSchneider98} were the first to suggest that strong
gravitational lenses could be used to detect dark matter substructure
through anomalous flux ratios of multiply imaged quasars.
Many groups subsequently followed up on this idea, both theoretically
and observationally
\citep{DalalKoch02,KochDalal04,Chiba05,SubaruMidIR,Macleod09,MaccioWDM,Chen07,Chen11,Rozo07,Keeton06,Chen03,Moustakas03,Metcalf04,Metcalf02,SchechWamb02,2016,Metcalf01,Keeton03,Fadely11,Keeton05,Xu09,Vegetti12}.
In particular, \citet{DalalKoch02} analyzed a sample of quadruply
imaged radio quasars from the CLASS survey \citep{CLASS}, and from the
preponderance of flux anomalies in these lenses, concluded that a
large fraction ($\sim 1-2\%$) of the projected mass at the Einstein
radius, $\sim 5$ kpc, must be in the form of local substructure.  The
uncertainties on this measurement were quite large, however, due to
the small sample size used in that analysis.  Subsequent work has not
significantly improved the bounds on substructure from strong lensing,
mainly due to the difficulties in using optical lenses to study
substructure \citep{KochDalal04}, however
\citet{Vegetti12} report a recent subhalo detection using an extended optical
galaxy-galaxy lens system.

Fortunately, in recent years a new class of lensed sources has
been detected: dusty, star-forming galaxies (DSFGs) at high redshift, $z\sim
2-5$ \citep{vieira:10, negrello:10}. 
DSFGs are a class of  luminous and prodigiously star-forming galaxies located at high redshift ($z>1$). 
They are enshrouded in dust \citep{blain02,lagache05} and contain massive reservoirs 
of molecular gas \citep{greve05,carilli:11}. The molecular gas in these galaxies is excited by
the intense high energy emission of the active star forming regions. 
Most commonly, CO, H$_2$O, HCN, and HCO+ are the molecular lines observed in these galaxies \citep{solomon05}.
In addition to molecular lines, atomic fine structure lines such as 
ionized carbon ([\cii]) has long been 
known to be a dominant cooling mechanism for DSFGs, and in some cases can 
account for $10^{-3}$ of their total infrared luminosity \citep{stacey91}. 
The intrinsic sizes and morphologies of DSFGs are not well understood, but
some studies \citep{chapman04,tacconi06, younger08, engel:10} suggest
typical radii of $\sim1$ kpc for these galaxies.
The submm emission from
DSFGs is believed to be dominated by  multiple compact clumps of intense star
formation spread out over the extent of the galaxy.
For example, \citet{swinbank:10} derive upper limits of $\sim$200 pc
for the diameters (FWHM) of star forming clumps in a lensed DSFG.

The brightest DSFGs were predicted to be strongly lensed
\citep{blain:96,negrello:07, hezaveh:11}.  This prediction was
confirmed by high-resolution follow-up imaging of the brightest extragalactic
sources found by the South Pole Telescope (SPT) \citep{Carlstrom:11,vieira:12, Hezaveh:12b}  and the \textit{Herschel Space Observatory} \citep{negrello:10}. 
Recently \citet{Hezaveh:12b} presented lens models for four  lensed DSFGs 
observed with ALMA in Cycle 0.  Even with very short total observing
times ($\sim20\, s$) and only $\sim15$ antennas, they were able to derive tight constraints on the mass
distributions of the foreground lenses, thanks to the extreme brightness of the sources
combined with ALMA's high sensitivity.  The stringent constraints that
are possible using snapshot observations with the small subset of
telescopes available in Cycle 0 strongly suggests that the full ALMA
array could provide an unprecedented view of the detailed mass
distributions of lensing galaxies.  In particular, ALMA observations of
lensed DSFGs could revolutionize the study of dark matter halo substructure
in lens galaxies.

Lensed DSFGs are a particularly attractive population for substructure studies, 
due to their great abundance (compared to quasar lenses).  In addition, 
since almost all of the UV and optical emission in the DSFGs is absorbed and reradiated
by dust at longer wavelengths, these sources are almost completely
invisible in optical images.  If substructure is detected
gravitationally, deep optical imaging would place stronger limits
on the mass to light ratios of the galactic satellites than would be 
possible if the source galaxies were bright in optical bands.
Moreover, the high redshifts of the sources permit a wide range of
possible lens redshifts,  potentially allowing constraints on 
any redshift evolution in the substructure population.

In this work we study the feasibility of using ALMA to detect dark matter
substructure in the halos of lens galaxies.
In particular, we study how the spatially resolved spectroscopy
provided by ALMA allows us to resolve source structures and dramatically increase
the substructure detection  
sensitivity. In \S\ref{sec:why} we discuss the benefits of
spectroscopically resolved interferometric observations of lensed
DSFGs.  In \S\ref{sec:simulation}  we describe our simulations
of ALMA observation of lensed DSFGs. In \S\ref{sec:analysis} and
\ref{sec:discuss} we present results of our simulations, and conclude
in \S\ref{sec:conclude}.

\section{Spatially Resolved Spectroscopy of Lensed Sources}
\label{sec:why}

One of the most important properties of lensed systems which
determines the sensitivity of observations to substructure lensing is
the size of the source.  A source that is completely uniform across
some scale $r_{\rm src}$ will be insensitive to lensing perturbations
from structures on scales small compared to $r_{\rm src}$.  When
studying dark matter substructure, it is therefore advantageous to use
lensed sources that intrinsically have structure on angular scales of
order milliarcseconds, comparable to the Einstein radii of the
subhalos of interest.  For example, radio quasars are excellent
sources for studying substructure, since their intrinsic sizes
($\theta \lesssim 0.1$ mas) are much larger than the Einstein radii of
individual stars but smaller than the Einstein radii of subhalos.  In
contrast, optical QSO's have much smaller angular sizes, meaning that
their lensed images can be affected by stellar microlensing which can
be difficult to disentangle from substructure lensing
\citep{KochDalal04,Morgan:12, Sluse:12}.   Similarly,
typical optically bright galaxies have much larger angular sizes, of
order $\sim 1\as$, rendering most galaxies insensitive to substructure
perturbations (however see \citet{Vegetti:09} and \citet{Vegetti12} for a 
method which uses the compact details of the structure in the source for 
substructure detection).

Naively, DSFGs would appear to be ill-suited as
sources for substructure lensing.  Typical DSFGs are believed to have
physical radii of order 1 kpc, corresponding to an angular scale of
order 0.2\as, much larger than the Einstein radii of all but the most
massive subhalos.  However, these galaxies do not have a smooth and
uniform morphology. They typically are believed to be composed of
multiple compact knots of star formation, based
both on theoretical 
and observational grounds \citep{swinbank:10}.  In principle, these
compact clumps should be sensitive to lensing perturbations on smaller
scales than the galaxy as a whole.  In practice, however,
the superposition of a large number of blended, overlapping
source clumps becomes indistinguishable from an extended source, thereby
reducing sensitivity to substructure lensing.  In order to fully
exploit the potential of lensed DSFGs as a probe of dark matter
substructure, we require some method to decompose the source emission
into its constituent compact clumps.

Fortunately, it is possible to perform such a decomposition in
velocity space.  The idea is analogous to the method of
\citet{Moustakas03} and \citet{Metcalf04}.
Spatially resolved spectroscopic observations of DSFGs
show strong velocity gradients, some indicating  fast-rotating disks,
and others indicating major mergers \citep{engel:10, riechers:11, Hodge:12}. If
the line-of-sight velocity offset between star forming clumps is larger
than the velocity dispersion of single clumps, then observations at
different frequencies will allow us to distinguish the emission from
distinct clumps.  In other words, the emission in a narrow frequency
window will come from a region in the source much smaller than the
overall extent of the DSFG.  Effectively, we can use 
{\em spectroscopic} resolution to enhance our {\em spatial} resolution.

Radio interferometers (such as ALMA) provide spatial and spectral resolution on interesting scales simultaneously, 
making these instruments
ideal observatories for probing substructure. We expect a
significantly enhanced sensitivity to substructure perturbations when
we simultaneously model the visibilities observed in all the channels,
compared to modeling the summed, channel-integrated visibility set.
In the next section, we demonstrate this enhanced sensitivity using
simulations of ALMA observations of lensed DSFGs.

\begin{figure*}
\begin{center}
\centering
\includegraphics[width=0.30\textwidth]{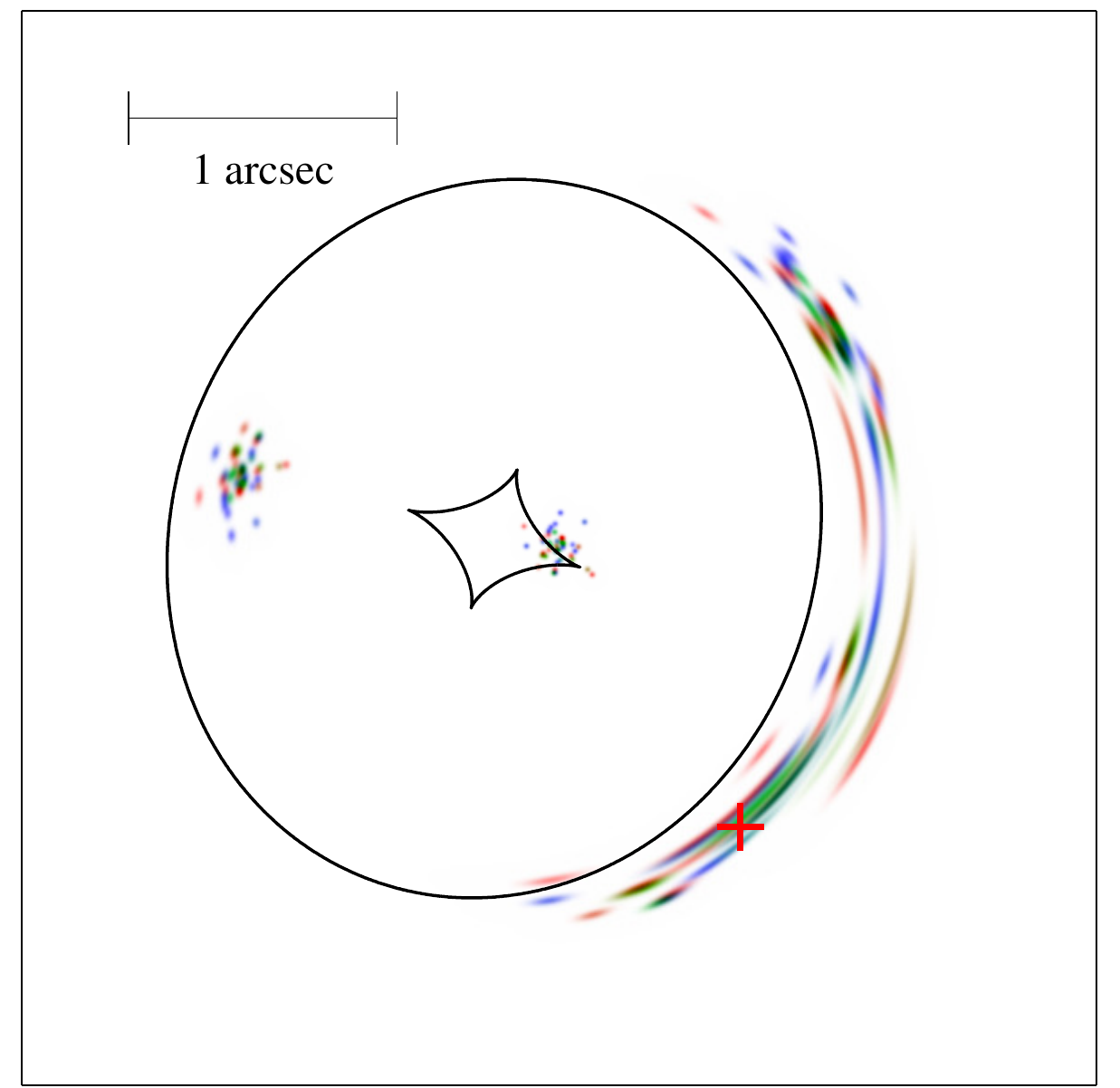}
\includegraphics[width=0.30\textwidth]{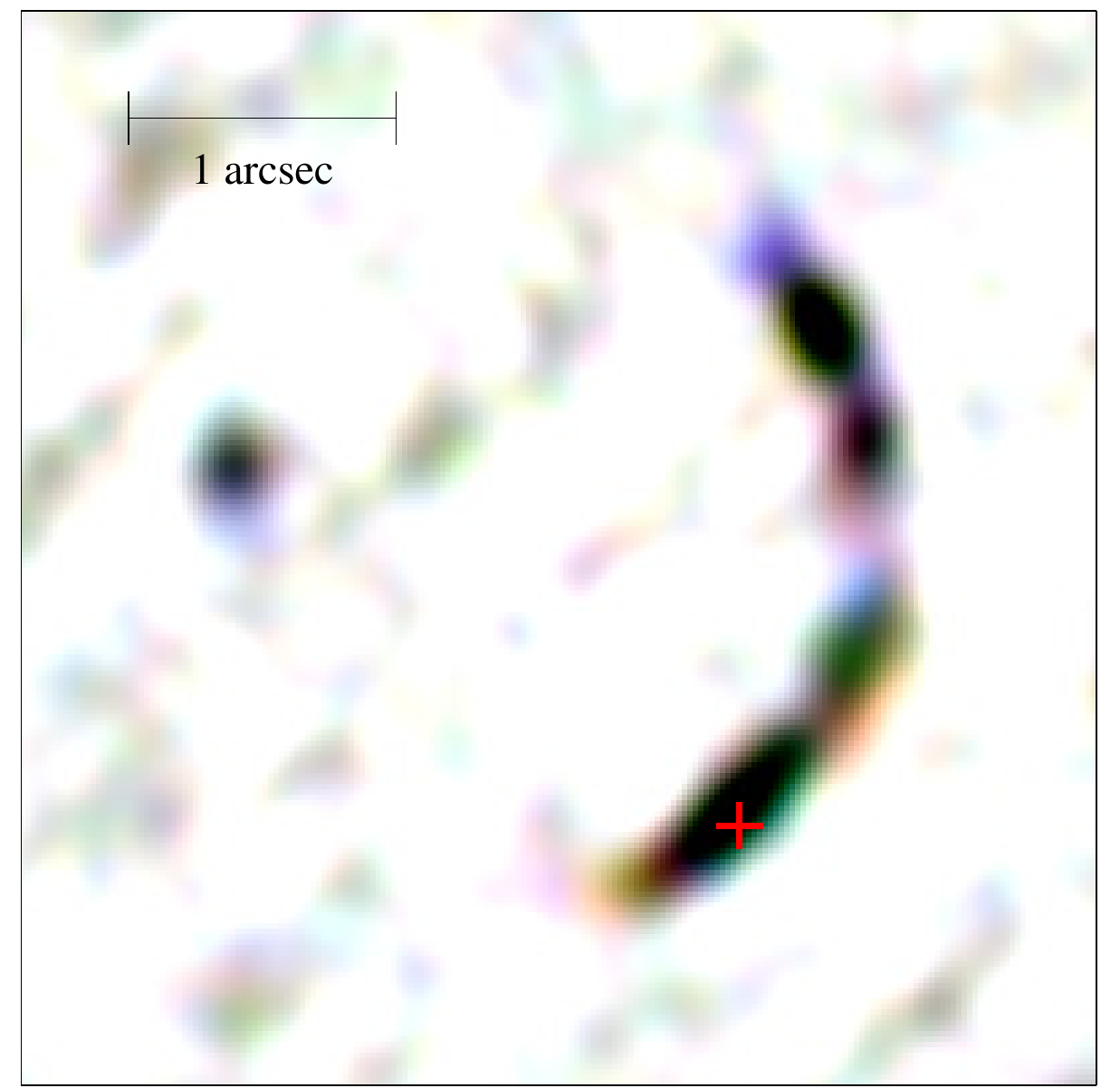}

\includegraphics[width=0.30\textwidth]{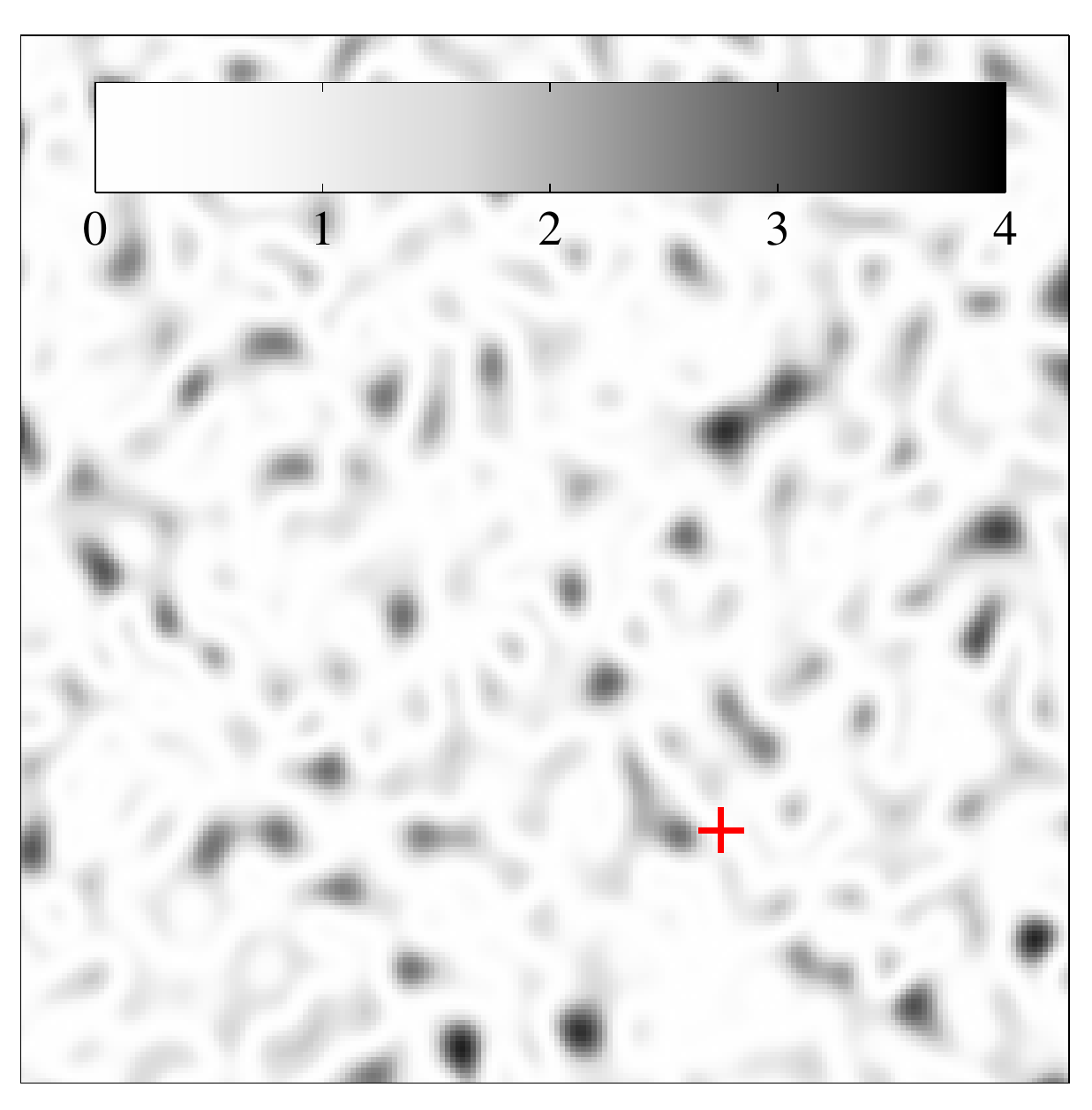}
\includegraphics[width=0.3\textwidth]{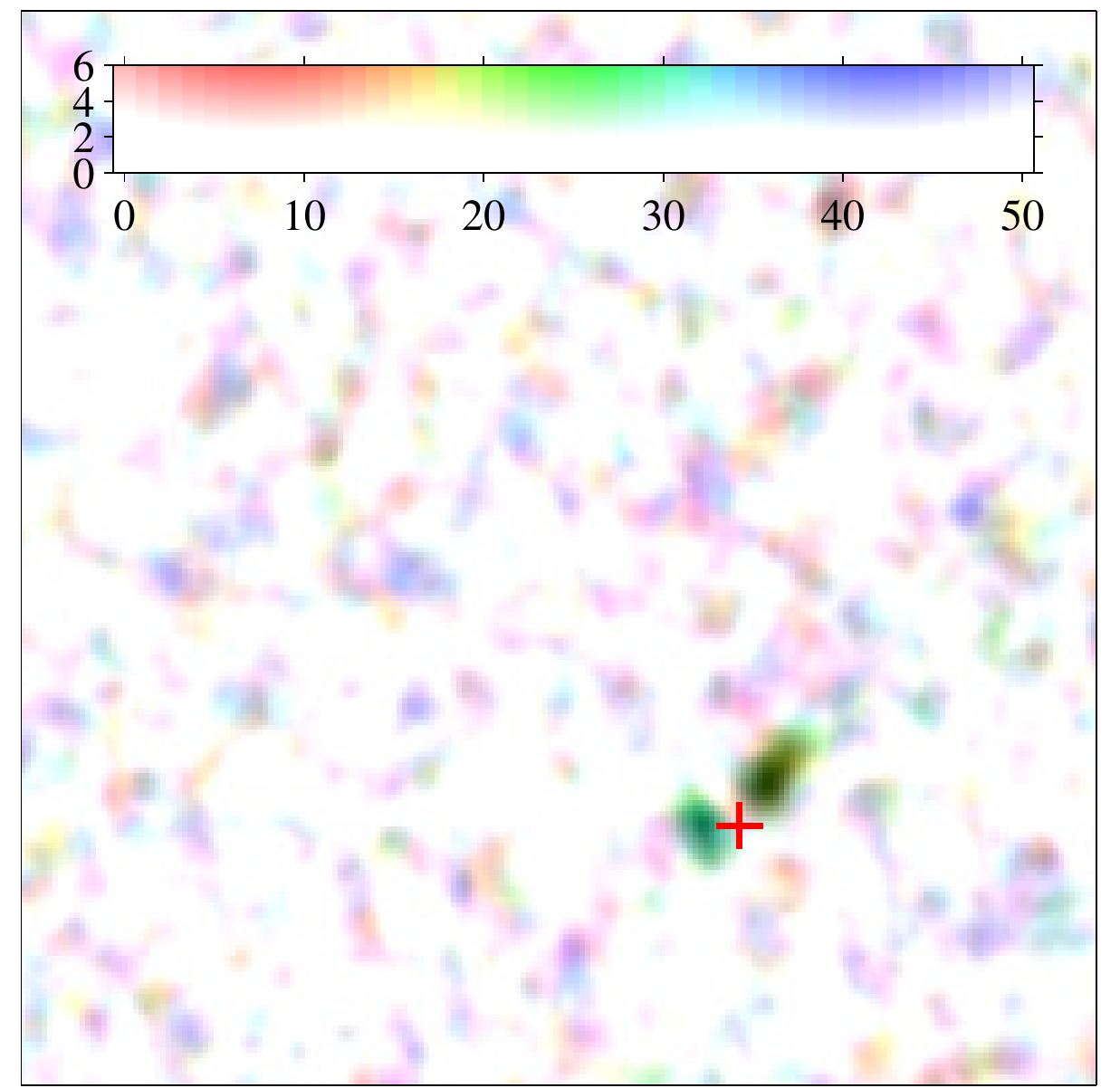}
\centering
\end{center}
\caption{ 
\textbf{Top Left:} The source model and the lensed images of a clumpy
source.  The black curves show the tangential and radial caustics for
the unperturbed macro model, and the red + symbol shows the location
of an additional subhalo of mass $M=10^8 M_\odot$.  
\textbf{Top Right:} The dirty image observed by ALMA.
\textbf{Bottom Left:} Residuals of channel integrated dirty images between a smooth model and noisy perturbed observation. The greyscale is in units of noise rms. 
 \textbf{Bottom Right:} Color residuals of the dirty images. The 50 observed channels are mapped to RGB colors as illustrated by the colorbar.   The y-axis of the colorbar shows the intensity in units of image noise rms in each channel.
\label{f:dirtyimage}}
\end{figure*}

\begin{figure*}[h]
\centering
\makebox[0cm]{\includegraphics[width=14.5cm]{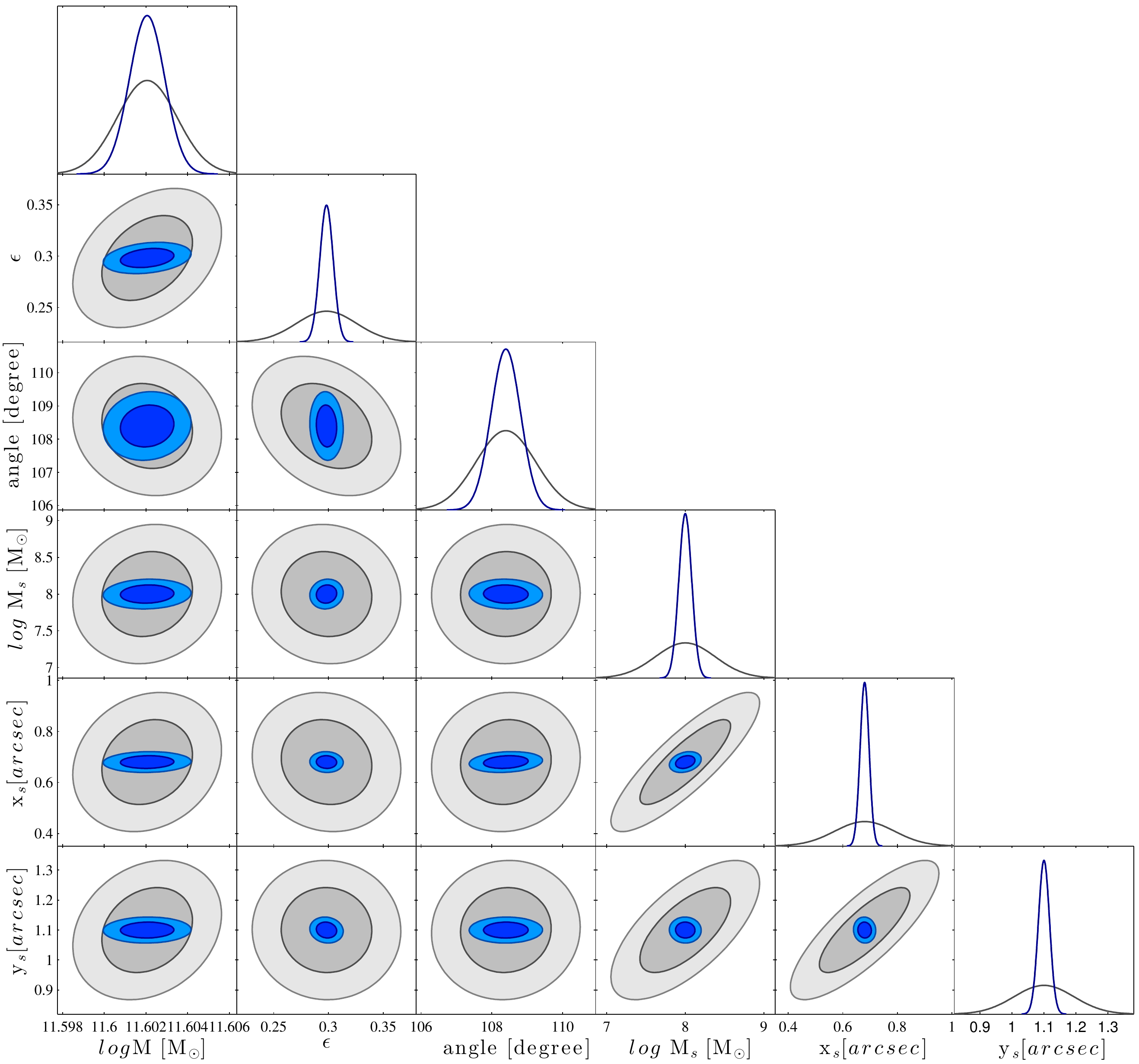}}
\caption{ Fisher matrix error forecasts. We simulate ALMA observations
  of a lensed DSFG, including substructure perturbations from a single
  subhalo of mass $M=10^8 M_\odot$ in the lens mass distribution, and
  compute the Fisher matrix for all the lens (including substructure) and source
  parameters.  We have marginalized over all parameters that are not
  shown in the figure. Gray contours correspond to fitting the
  channel-integrated flux, and blue contours correspond to fitting all
  channels simultaneously.} 
\label{f:fisher}
\end{figure*}

\section{Simulations}
\label{sec:simulation}
We simulate observations of lensed DSFGs with ALMA Cycle 1 (32
antennas) with the array in its most extended configuration, with a
maximum baseline of 1.1 km. This allows a resolution of $0.16''$ at
350 GHz.  The noise levels are calculated using the online ALMA
Sensitivity Calculator for best observing conditions ($1^{st}$
Octile: PWV=0.5 mm, $T_{sys}$=90~K).  Note that the full ALMA
array will have roughly 4 times greater sensitivity, and more than 10
times higher spatial resolution than what we have assumed in the calculations
presented here.  We place the source at Dec -50:30 and simulate a one
hour observation with 10 second integration time. The uv-coverage was
predicted using the ``simdata'' task of the Common Astronomy Software
Application (CASA) package \citep{Petry:12}.    

 The source is modelled as a collection of giant star-forming clumps.
 Since we focus on high-excitation lines that are found only in high
 density clumps, we will neglect any diffuse emission originating from
 the interclump ISM.  In local starburst galaxies, and in high redshift DSFGs, 
 CO line brightness peaks at $J = 6$, and at higher $J$ in AGN-dominated sources \citep{Weiss:07, Lestrade:10, Rangwala:11}. Since transitions such as CO 6-5 produce
some of the brightest lines and trace the dense cores of star forming clumps  
with compact morphologies, they may be the ideal molecular lines for
substructure lensing detections. 

The clumps are placed randomly in both spatial location and velocity,
with positions drawn from a gaussian profile with rms of 1 kpc and
velocities drawn from a uniform distribution of width 300 km/s. Each
clump has a circular Gaussian surface brightness profile with FWHM of
$d_c=250$ pc, and a Gaussian profile in redshift space characterized
by a velocity dispersion of $\sigma_v=30$ km/s. For simplicity we have
used random velocities for the clumps, as opposed to a rotating disk,
for example.  As long as the clumps are reasonably well separated in
velocity space ($\Delta v \gtrsim \sigma_v$) we do not expect any
qualitative differences between ordered motion and disordered motion.
Each source clump is therefore described by 6 parameters: $x$ and $y$
centroids, central velocity, radius, flux and velocity dispersion.
The source emission, summing over all clumps, is represented as a 3D
data cube. The total intrinsic velocity integrated flux of the entire
DSFG is set to 1 Jy km/s (see \citet{bothwell:12} for similar values
for the velocity integrated CO flux of unlensed DSFGs) and
distributed equally among  10 star forming clumps.

Given our source realizations, we then ray-trace to compute a new 3D
data cube containing the lensed images observed in each channel.  We
then Fourier transform each layer of the cube, and predict the
visibilities by interpolating the Fourier maps.  

In each simulation the main lens galaxy is modelled as a Singular
Isothermal Ellipsoid (SIE), which has a three dimensional mass density
proportional to $r^{-2}$, and projected surface density
\begin{equation}
\Sigma(x,y) = \frac{\sqrt{q}v^2}{2G}\left(\frac{x^2+q^2 y^2}{r_{\rm E}^2}\right)^{-1/2}
\end{equation}
where $x$ and $y$ are coordinates oriented along the principal
axes of the surface density measured relative to the lens centroid,
$v$ is the velocity dispersion along the line of sight, $q$ is the
axis ratio, and $r_{\rm E}$ is the Einstein radius of the lens,
\begin{equation}
r_{\rm E} = 4 \, \pi \frac{v^2}{c^2} \, \frac{D_d \, D_{ds}}{D_s}
\end{equation}
\citep{Kormann:94}.  The SIE therefore is described by 5 parameters:
the centroid $x_c$, $y_c$, velocity dispersion $v$, axis ratio $q$,
and orientation $\theta_q$.  
We simulate lenses with Einstein radius mass
$M_{\rm E} = \pi \Sigma_{cr} r_{\rm E}^2$ of $4\times 10^{11} M_{\odot}$ 
placed at $z_d=0.5$ and place the source at $z=2$. 
In addition to this main lens, we allow for
external shear, described by an amplitude $|\gamma|$ and orientation
$\theta_\gamma$.  The smooth mass model therefore has 7 parameters
describing the lens.  
We follow previous work and model dark matter subhalos using the Pseudo-Jaffe
density profile \citep{Munoz01}, 
\begin{equation}
\kappa(x) =  \frac{1}{2x} - \frac{1}{2\sqrt{x^2+x_t^2}}~,
\end{equation}
where $x=r/r_{\rm E}$ and $x_t = r_t/r_{\rm E}$ are the radius and the
tidal truncation radius, respectively, in units of the subhalo's Einstein
radius, and $\kappa = \Sigma(r) / \Sigma_{cr}$ is the dimensionless
surface density expressed in units of the lensing critical surface
density \citep{Schneider92}.

Figures \ref{f:dirtyimage} and \ref{f:fisher} illustrate the increased
sensitivity to substructure provided by spectroscopically resolved
data. Figure~\ref{f:dirtyimage} shows an example of ALMA observations
of a lensed DSFG, both for spectroscopically resolved visibilities and
channel-integrated visibilities.  The greyscale and
colored panels show the residuals from the best-fitting smooth lens
models for each case, and the substructure perturbation clearly stands
out more readily in velocity space. In Figure~\ref{f:fisher}, we show
the substructure parameter errors derived from simulated observations of another
lensed DSFG, whose properties were chosen to be representative of the
lenses found in \citet{Hezaveh:12b}. The parameter uncertainties
plotted in this Figure 
were estimated by a Fisher matrix calculation; we
marginalize over a considerably larger number of nuisance parameters
describing the source emission when fitting spectroscopically resolved
visibilities, compared to fitting the channel-integrated visibilities.
Despite the increased number of marginalized nuisance parameters, the
parameter uncertainties are considerably reduced when we utilize the
full, velocity-resolved data cube (compare grey vs.\ blue contours).
This is true for the ``macro model'' parameters describing the smooth
lens, and for the subhalo parameters as well.

ALMA observations will therefore clearly improve our ability to
characterize substructure in strong lenses.  
As Figure~\ref{f:fisher} demonstrates, the properties
of a $M=10^8 M_\odot$ subhalo  can be
determined with high precision, especially when
velocity information is utilized in the model fits. 
However, before undertaking a
systematic study of subhalo parameter measurements,
it is important to estimate the substructure detection rates
expected in ALMA observations of typical lens systems.  Will detections 
like the subhalo shown in Fig.~\ref{f:fisher} (a $6.8\sigma$
detection) be common or rare in DSFG lenses?
The focus of this paper is to demonstrate the benefits of velocity fitting,
and to quantify the expected substructure detection rates using velocity fitting with ALMA.

We define the significance of a subhalo detection in terms of the
residual $\chi^2$ between the best-fit smooth mass model and the
simulated observations.
In our simulations we start with a smooth macro model and add
deflections from a single subhalo during the ray-trace.  We then model
the simulated observations using only the smooth mass model, with no
substructure parameters, allowing the macro model parameters to adjust
in order to account for the substructure perturbations.

\begin{figure*}
\begin{center}
\centering
\includegraphics[width=0.24\textwidth]{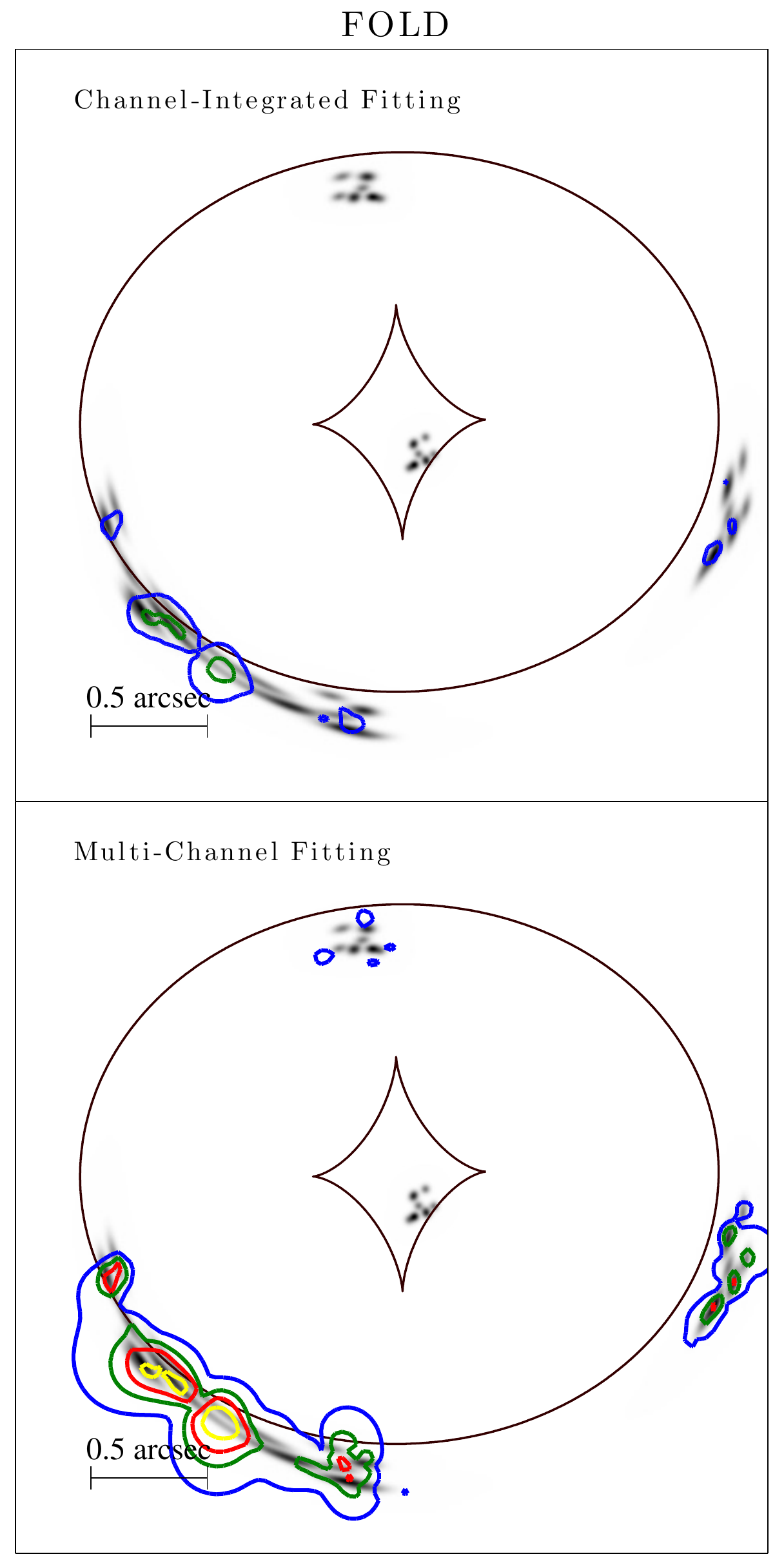}
\includegraphics[width=0.24\textwidth]{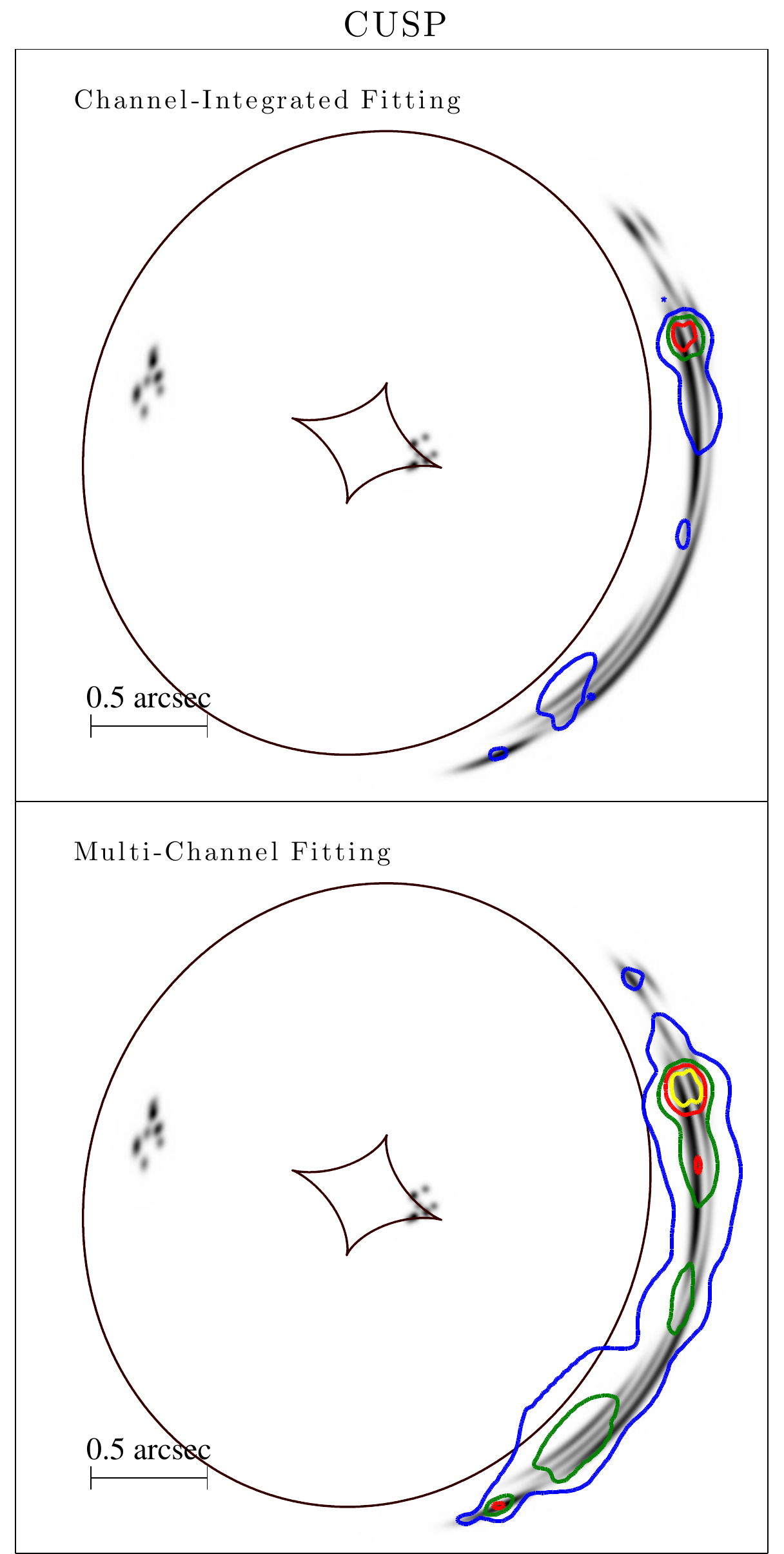}
\includegraphics[width=0.24\textwidth]{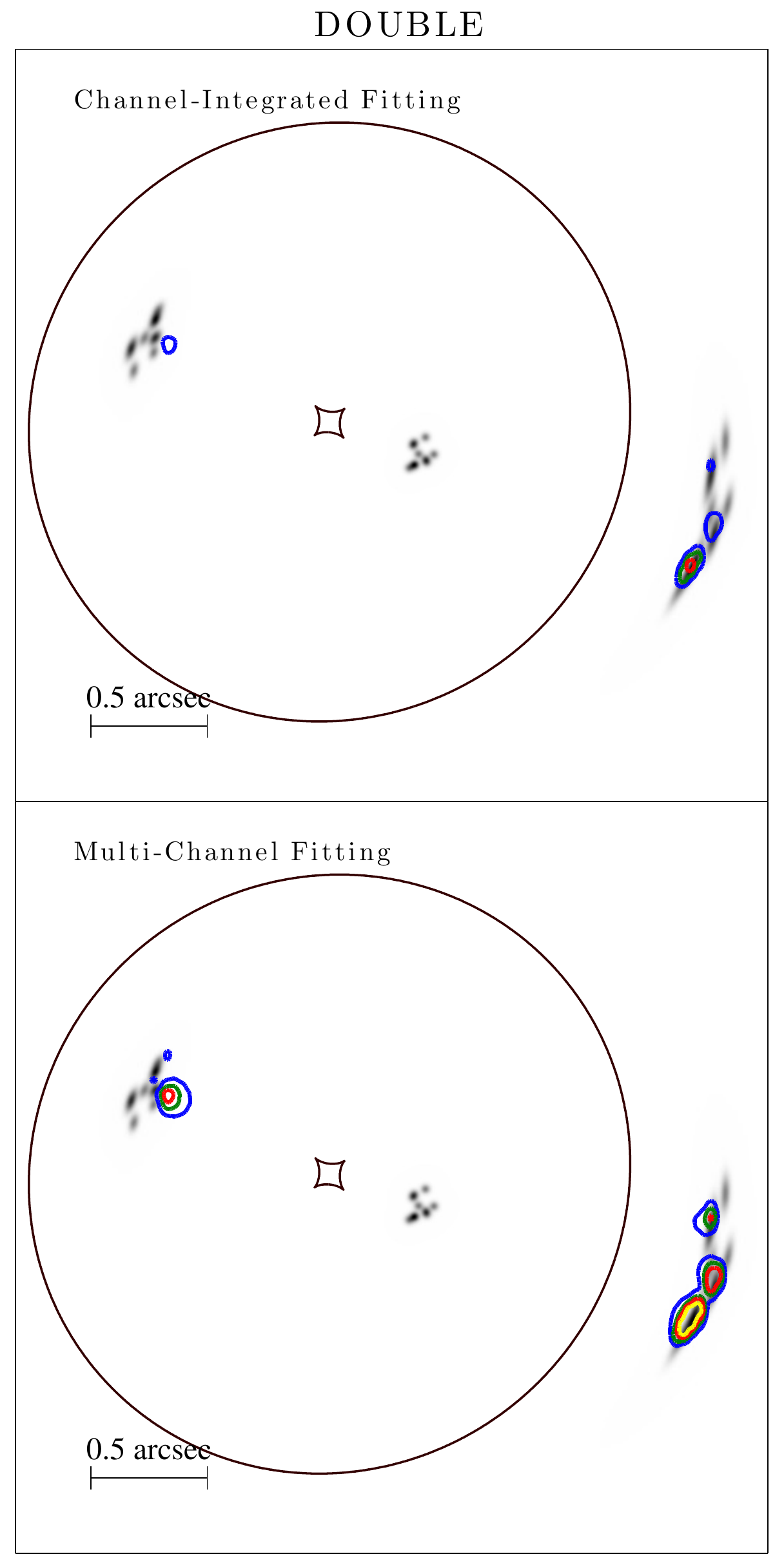}
\includegraphics[width=0.24\textwidth]{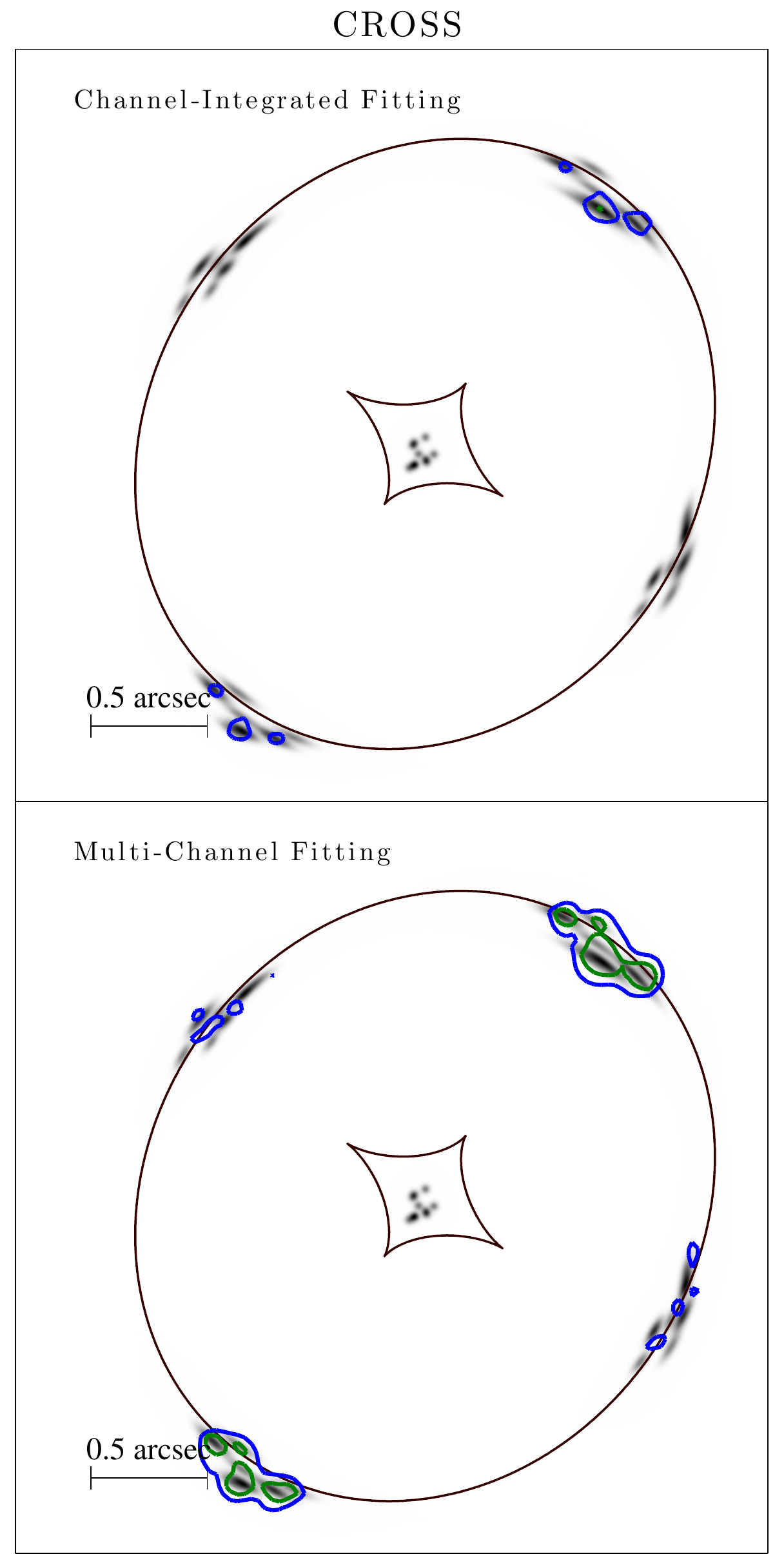}

\centering
\end{center}
\caption{ 
Substructure detection cross-sections.
Contours show the detection significance for substructure as a
function of position for a subhalo of mass $M=10^8 M_\odot$.  The
different colors correspond to  3 (blue), 5 (green), 7 (red), and 10
(yellow) sigma  detections.  
Columns correspond to different lensing configurations labeled on top
(fold, cusp double, and cross from left to right). The greyscale image
shows the model lensed images. The unlensed source is also plotted at
the center. The black curves show the tangential and radial caustics,
not the critical curves. The bottom row shows the detection
significance when fitting to all channels simultaneously, while the top
row shows detection significances for channel integrated visibilities.  
 \label{f:contours}}
\end{figure*}

\begin{figure}[h]
\begin{center}
\centering
\includegraphics[width=0.45\textwidth]{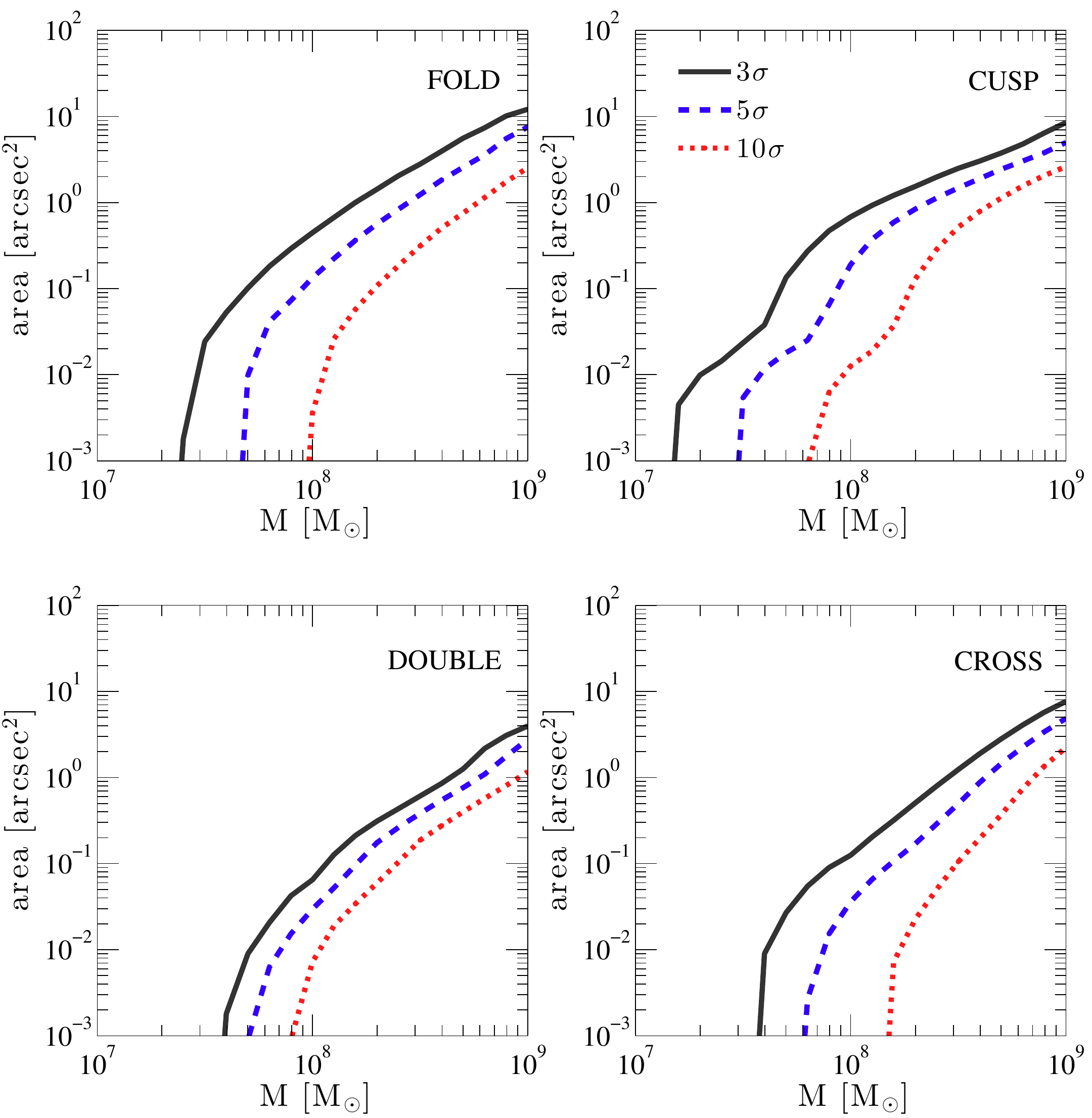}
\caption{Detection cross-section for subhalos as a function of mass.
  Each panel corresponds to a different macro lens image
  configuration. The curves show the area of 3 (solid black), 5 (dashed
  blue), and 10 (dotted red) $\sigma$ detection significance.
\label{f:area}
 }
\centering
\end{center}
\end{figure}

\begin{figure}[h]
\begin{center}
\centering
\includegraphics[width=0.45\textwidth]{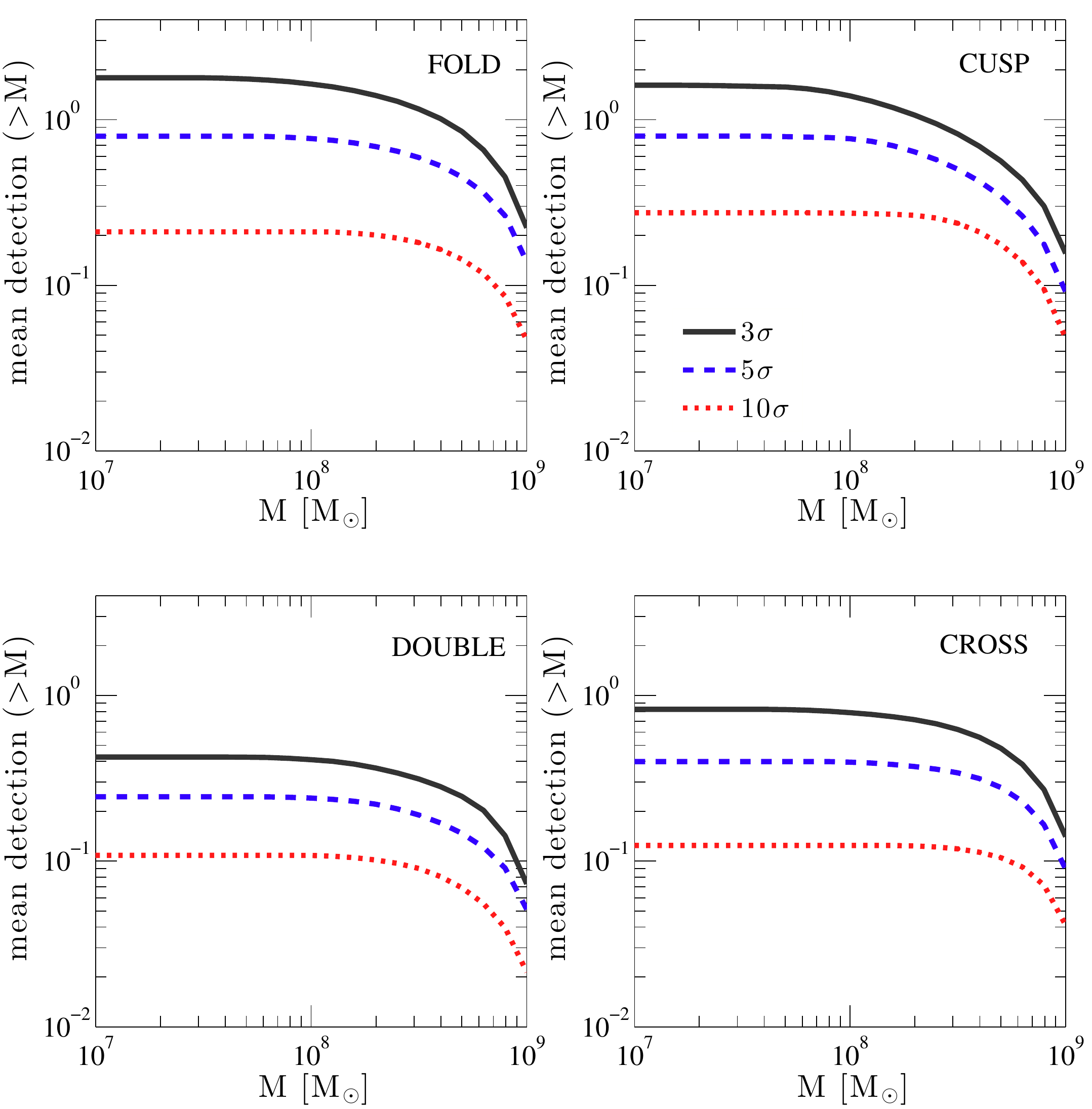}
\caption{Left: Cumulative mean number of detections of subhalos
  of mass greater than M. The curves show the number of detections
  with 3 (solid black), 5 (dashed blue), and 10 (dotted red) sigma
  significance. \label{f:numdetections}
}
\centering
\end{center}
\end{figure}

We follow \citet{DalalKoch02} and use linear perturbation theory to
determine the parameters of the best-fitting smooth macro model.
We characterize the smooth macro model with a parameter set $\bm{p}$
(including the source profile parameters). For some fiducial set of
parameters $\bm{p}_0$, we obtain observables $\bm{O}$ (a vector of
real and imaginary visibilities) and write the penalty function  
\begin{equation}
\chi^2 = \delta O_i(\mathbf{C}^{-1})_{ij} \delta O_j
\end{equation}
where summation over repeated indices is understood, and $\mathbf{C}$
is the noise covariance matrix.  If our observables 
are the visibilities, then {\bf C} is diagonal.

A perturbation to the model parameters $\delta\bm{p}$ generates
perturbations to the observables given by $\delta O_i = (\partial O_i/\partial p_j) \delta
p_j$. Given a current data residual $\delta \bm{O}$, the addition of
these parameter adjustments changes the $\chi^2$ to
\begin{equation}
\chi^2 =  \left[ \delta O_i  + \delta p_k\frac{\partial O_i}{\partial p_k} \right]       (\mathbf{C}^{-1})_{ij}         \left[ \delta O_j  + \frac{\partial O_j}{\partial p_l}  \delta p_l \right] 
\end{equation}

To find the parameter set $\bm{p}$ that minimizes this function we set
$\partial \chi^2/\partial p_i=0$ and solve for $\delta p_i$ 
\begin{equation}
\delta p_l = -(\mathbf{F}^{-1})_{lk} \frac{\partial O_j}{\partial p_k} (\mathbf{C}^{-1})_{ji} \delta O_i
\label{par:adj}
\end{equation}
where
\begin{equation}
F_{ij}  = \frac{\partial O_k}{\partial p_i} (\mathbf{C}^{-1})_{kl}   \frac{\partial O_l}{\partial p_j} 
\label{eq:fisher}
\end{equation}
is the Fisher matrix for the smooth model. 

Using this procedure we choose a smooth macro model and calculate the
matrix  $\partial O / \partial p$ by finite differencing. This
matrix has dimensions $n_{\rm obs}\times n_{\rm par}$, where $n_{\rm
  obs}=18$ million is the number of observables (real and imaginary visibilities in 50 channels for a 1 hour observation), and
$n_{\rm par}=57$ is the number of source parameters (7 for the lens
model, and 5 parameters each\footnote{We did not vary each clump's
  velocity dispersion $\sigma$, since we did not expect it to be
  degenerate with any parameters of the lens mass model.}
 for 10 independent source clumps).  
We use this matrix to calculate the Fisher matrix {\bf F} using
equation (\ref{eq:fisher}).  We add a  single subhalo to the current
smooth model, predict the new visibilities, and use  
equation (\ref{par:adj}) to find the parameter adjustments to the smooth
model that minimize $\chi^2$.  We then simulate a smooth
macro lens observation with parameters $\bm{p}+\delta\bm{p}$ and use the new
$\chi^2$ between this best-fit model and the original simulated
observation as the likelihood of a detection. 

We simulate four different lensing configurations labeled
``fold'', ``cusp'', ``double'', and ``cross''.  All configurations have
parameters that are typical for strong lensing systems.  
As before, the background source is chosen to consist of 10 Gaussian clumps, of diameter (FWHM)
250pc and velocity dispersion of 30  km/s.  These values are
chosen to be consistent with well-resolved observations of molecular gas clumps in lensed DSFGs \citep{swinbank:10,swinbank:11}.  Our sensitivity to substructure
depends strongly on the sizes of these clumps, as discussed in
\S\ref{sec:discuss}.  For each configuration we define
a Cartesian grid of substructure positions.  We place a single subhalo
at each location and use the above procedure to measure the detection
significance of substructure with mass M as a function of position.
The area inside a contour of specified $\chi^2$ then defines the
detection cross-section for each significance level.

We perform simulations for 20 different substructure masses, between
$10^7  - 10^9 M_{\odot}$.   Our results, shown in Figures \ref{f:area}
and \ref{f:numdetections}, are discussed in the next section.

\section{Results}
\label{sec:analysis}

We define the substructure detection cross-section as the area of the
sky (e.g. in square arcsec) inside of which a subhalo can be
detected with a given minimum significance. We compute this
cross-section by calculating the detection significance, defined in
\S\ref{sec:simulation}, as a function of subhalo position.  Figure
\ref{f:contours} shows the detection cross-sections at 3, 5, 7, and 10
sigma significance for a $M=10^8 M_{\odot}$ subhalo, for four
different macro lens configurations.  The top row shows this
cross-section when the channels are integrated before fitting and the
bottom panel shows the cross-section when fitting to each of the
channels individually. The different extent of the detection
cross-section for different lensing configurations suggests that there
is a higher probability of detecting sub-halos in high-magnification
fold and cusp image configurations compared to low-magnification cross
or double configurations.  Nonetheless, the low-magnification lenses
retain some sensitivity to substructure.  This is unlike the case of
2-image quasar lenses, which generally lack sufficient constraints to
permit substructure detection.  For DSFG lenses, the presence of
multiple source clumps provides enough constraints on the mass model
to permit substructure detection in favorable configurations.

We perform similar simulations for other subhalo masses ranging
between $10^7 M_{\odot}$ to $10^9 M_{\odot}$, and calculate the total
cross-section areas for each mass bin. Figure \ref{f:area} shows these
cross-sections as a function of mass for each lensing configuration.
We have only plotted the cross-sections derived from channel-fitting,
having already established that the detection sensitivity will be
considerably diminished for channel-integrated data.

These cross-sections, in combination with a substructure mass
function, can be used to estimate the expected number of detected
subhalos for each DSFG lens.  We assume a subhalo mass function with a
slope  of $d\log N/d\log M=-1.9$, an upper mass bound of $10^9 M_{\odot}$ and an overall normalization placed by 
setting the mass fraction of substructure inside the Einstein radius
of the macro lens to $f=1\%$, consistent with results of the quasar
lensing analysis of \citet{DalalKoch02}.  
Multiplying the sky number density of subhalos with masses
between $M$ and $M+\delta M$ by the detection cross-section of mass
$M$ gives the average number of detections with the given
significance, assuming a uniform distribution of substructure over the
strong lensing region. Figure \ref{f:numdetections} shows the
cumulative number of detections for the four lensing configurations.
Note that configurations with high magnification (the fold and cusp
configurations) give more than 1 detected subhalo on average (at
$3\,\sigma$ confidence) for each lensing system.  This somewhat
invalidates our treatment, which explicitly 
assumes that only a single subhalo provides perturbations.  In
contrast, the collective perturbations of multiple subhalos must be
simultaneously treated.  In forthcoming work, we will analyze the
perturbations from the population of substructure inside dark
matter halos.

\begin{figure}[h]
\begin{center}
\centering
\includegraphics[width=0.48\textwidth]{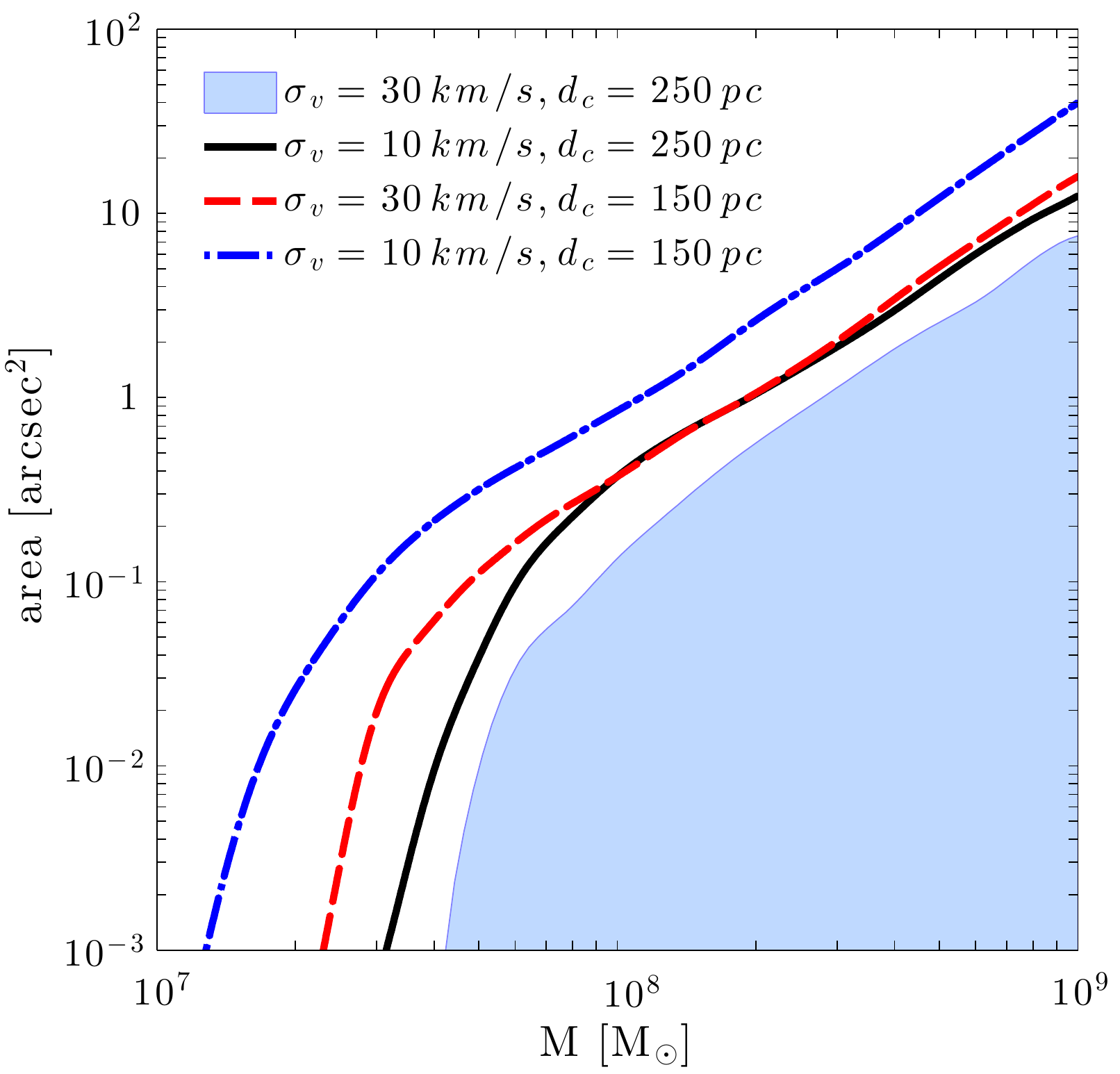}
\includegraphics[width=0.48\textwidth]{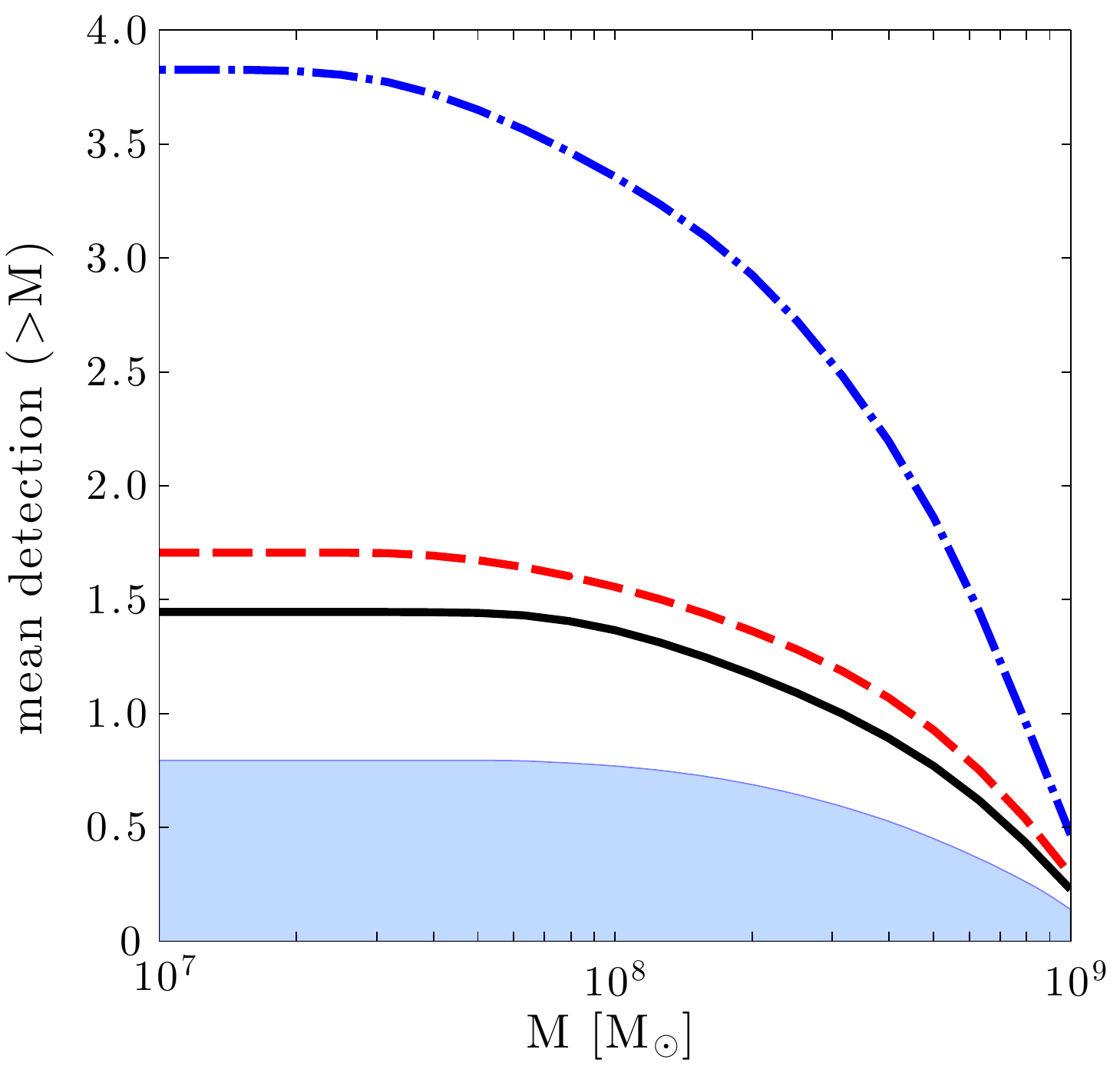}
\caption{Detection sensitivity as a function of source properties.
  Here we show the improvement in detection significance when observing
  DSFGs composed of smaller clumps (red dashed curves), clumps with
  lower velocity dispersion (solid black curves), or both (dot-dashed
  blue curves), compared to our fiducial source parameters (blue
  shaded region), for the fold configuration. The top panel shows the
  detection cross-section, and the bottom panel shows the mean
  cumulative number of detections.  All curves are shown at 5 $\sigma$
  detection significance. Smaller clumps could correspond to observing
  dense-gas tracers such as HCN. \label{f:HCN}
}
\centering
\end{center}
\end{figure}

\section{discussion}
\label{sec:discuss}

In the simulations presented in this work, we have assumed a fixed
source morphology with reasonably conservative parameters describing
the DSFGs.  In this section, we briefly explore the sensitivity of our
results to the assumed properties of the source clumps.  

As discussed above, one of the most important parameters describing
the source is its angular size.  In the calculations we have presented
so far, we have assumed a physical size of 250 pc (FWHM) for the
source clumps.  We chose this value as a conservative upper bound,
given the results of \citet{swinbank:10}.  The CO-emitting regions of
the star forming clumps could be significantly smaller than this
assumed size, however.  Upcoming ALMA observations of DSFGs should
resolve the question of the intrinsic source size.  Even if the source
clumps are not significantly smaller than our assumed values,
observations of other molecular lines besides CO lines could reveal much smaller
source sizes and different source morphologies
\citep{carilli:11,riechers:11,Combes:12}.
Transitions of H$_2$O, HCN, and HCO$^+$ have higher critical densities than CO at similar observing frequencies. These lines trace dense gas in the active star forming regions \citep{Downes:98}, and may have fluxes as high as 25\% of CO flux
\citep{Gao:04} or more \citep{riechers:10}. Because they are confined to smaller regions than low and possibly high-$J$ CO, they may increase the sensitivity to lower mass subhalos.
To demonstrate this enhanced sensitivity, we have repeated
the fold configuration simulations presented in Figure \ref{f:area}
with identical parameters, but using smaller source clumps (150 pc
FWHM). The red dashed curve in the top panel of Figure \ref{f:HCN}
shows the $5 \, \sigma$ detection cross-sections of subhalos with
masses ranging from $10^7$ to $10^9 \, M_{\odot}$. The blue shade
corresponds to the larger clump sizes presented in Figure
\ref{f:area}. The bottom panel shows the resulting increase in the
mean number of detected subhalos, demonstrating that different
molecular lines originating from different morphological structures of
varying sizes exhibit different levels of sensitivity to substructure.

Just as a smaller angular size improves the sensitivity to
substructure, narrower linewidths also help resolve the emission from
distinct clumps.  
If the intrinsic velocity dispersion of each star forming clump is
lowered, then fewer clumps appear simultaneously in each channel.
This effectively shrinks the angular extent of the emitting region in
each channel, increasing the substructure detection sensitivity. In
Figure \ref{f:HCN} the black solid curves show the $5\, \sigma$
detection cross-section (top) and mean cumulative detection numbers
(bottom) of a DSFG composed of clumps with velocity dispersion of 10
km/s \citep{Hopkins:12}. The key factor here is the relative velocity
offset of clumps with respect to each other, in units of their
velocity dispersion.  The blue dot-dashed curve in Figure \ref{f:HCN}
shows the detection improvements for simulations with both smaller
clumps and lower velocity dispersion.  As the bottom panel of Figure
\ref{f:HCN} illustrates, for a source with these parameters we expect
to detect more than one subhalo in a given lens, if the substructure
abundance is consistent with previous lensing analyses.

Figure \ref{f:HCN} demonstrates that our sensitivity to substructure
depends on the clump morphology, both in real space and in velocity
space.  We have chosen fiducial values for these source parameters to
be consistent with current measurements of the intrinsic properties of
DSFGs. 
The unlensed DSFGs presented in \citet{engel:10} either showed
signatures of merging systems, or fast rotating compact
morphologies. Comparing the values of
galaxy-scale velocity dispersions reported in \citet{engel:10}  which
range between 200 and 1000 km/s (FWHM), with the $\sim$ 10 km/s
dispersions expected for single clumps \citep{Hopkins:12}, indicates
that all the DSFGs in their sample must contain several components
with significant velocity offsets.  Similarly, the double peak velocity profile 
presented in \citet{Hodge:12} is indicative of large velocity offsets 
between multiple components. This suggests that systems with
very small velocity offsets between their multiple components may be
rare, so our assumption of $\Delta v \gtrsim \sigma_v$ appears to be
safe, even in the case of rotationally supported DSFGs.

Our fiducial simulations used a fixed number of 10 source clumps,
randomly distributed across a galaxy of size 1 kpc.  Our results do
not appear to be as sensitive to these parameters as the clump size
or linewidth, although this may depend on the details of the
image morphology.  For example, increasing the galaxy size
allows the clumps to cover a larger portion of the caustic, which
tightens the constraints on the macro model.  However, it also
increases the spacing between clumps, thereby degrading the
sensitivity to low-mass substructure.  The effect of changing the
number of clumps is also somewhat unclear.  In principle, a larger
number of clumps should provide a larger number of constraints.
However, since we hold fixed the total flux, increasing the clump
number makes each clump fainter and lowers their signal to noise
ratio.  We found that for the fold configuration described above,
decreasing the number of clumps from 10 to 5 somewhat improves the
detection sensitivity, mainly by making each clump brighter.  It is
unclear whether this will hold for the other configurations as well.
In the absence of a systematic study all possible image
configurations, we cannot say with any certainty how our substructure
detection sensitivity will depend on the number and spread of the
source clumps.

Although in this work we have focused on high excitation molecular
lines emitted by the compact cores of star forming clumps, the diffuse
emission in low-excitation lines may also benefit from spatially
resolved spectroscopy.  Rotationally supported cold disks exhibit strong
velocity gradients on the sky, meaning that the emission in narrow
channels will originate from regions significantly smaller than the
galaxy as a whole.  In such cases the observed velocity gradients in
the lensed images of fast rotating cold gas reservoirs may show dips
and peaks consistent with substructure lensing, complementing the
high-$J$ line observations.   
 
The calculations presented here were based on simulations of 1-hour
observations using ALMA Cycle 1.  Longer observing times would improve
our quoted signal/noise ratios by $t^{1/2}$.  In addition, the
simulations presented here were specifically carried out for early
science capabilities of ALMA with an array of only 32 antennas and a
maximum baseline of $\sim$1 km. The full ALMA array will consist of 64
antennas, which will increase the sensitivity by a factor of 4. In
addition the longest baselines will extend as long as $\sim$16 km,
resulting in an angular resolution of $\sim$10 milliarcsec at 850
$\micron$.  We can state with confidence that the full ALMA array will
provide far greater substructure sensitivity than our Cycle 1
calculations have already found.

In this work we have focused on ALMA observations of DSFGs, but in
principle, the methods used here are not limited to interferometric
observations.  For example, integral field spectroscopy
\citep[e.g.][]{Alaghband-Zadeh:12,Barnabe:11,FoersterSchreiber} of
other lensed galaxies such Lyman Break Galaxies (LBG) may benefit in
the same way.   

\section{Conclusion}
\label{sec:conclude}

We have shown that it is possible to detect galactic dark matter
substructure in the lens galaxies of lensed DSFG systems, using the
early science capabilities of ALMA in Cycle 1.
We found that the analysis of spatially resolved spectroscopic
measurements of lensed sources can significantly improve substructure
detection limits. In particular we simulated ALMA Cycle 1 observations
of molecular lines in these systems, and calculated the detection
significance for subhalos of different masses for various lensing
configurations.  We predict that with current ALMA capabilities and
conservative assumptions about the morphology of DSFGs, there is a
high probability of detecting at least one subhalo with $M \gtrsim
10^8 M_{\odot}$ in {\em every lensed DSFG}.  Only
marginally more optimistic source morphologies allow us to go below
this limit and explore $M\sim10^7 M_{\odot}$ dark matter subhalos.
The full ALMA array, with $4\times$ greater sensitivity and $16\times$
higher angular resolution than what we have assumed, will assuredly
improve our sensitivity to substructure beyond the calculations
presented here.  Given samples of $\sim$100 lensed DSFGs
discovered in large area millimeter surveys and the significantly
enhanced capabilities of the completed ALMA, we are poised to
revolutionize our understanding of low-mass dark matter substructure in
coming years.

\acknowledgements{ This work was supported by NASA under grant
  NNX12AD02G,  by NSERC, the CRC
  program, and CIfAR. YDH is grateful for useful discussions with C. Fassnacht 
  which considerably improved the uv-plane modeling.
  YDH acknowledges the support of FQRNT through
  International Training Program and Doctoral Research scholarships.
  ND gratefully acknowledges support by a Sloan Research Fellowship
  from the Alfred P.\ Sloan Foundation. We thank the SPT team 
  for numerous useful discussions.
}

\bibliographystyle{apj}
\bibliography{references}

\end{document}